\documentstyle[12pt,epsfig]{article}
\begin{document}
\thispagestyle{empty}
\begin{titlepage}
\begin{center}
\hfill FTUV/98-67\\
\hfill IFIC/98-68\\
\vskip 1.5cm
\LARGE
Generalized Hypergeometric Functions and the Evaluation of
Scalar One-loop Integrals in Feynman Diagrams
\end{center}
\normalsize
\vskip1cm
\begin{center}
{\large \bf Luis G. Cabral-Rosetti}\footnote{E-mail: cabral@titan.ific.uv.es}
{\large and}
{\large \bf Miguel A. Sanchis-Lozano}\footnote{E-mail: mas@evalo1.ific.uv.es}
\end{center}
\begin{center}
\baselineskip=13pt
{\large \it Departamento de F\'{\i}sica Te\'orica and IFIC\\
Centro Mixto Universidad de Valencia-CSIC}\\
\baselineskip=12pt
{\it 46100 Burjassot, Valencia (Spain)}\\
\vglue 0.8cm
\end{center}
 
\begin{abstract}

Present and future high-precision tests of the Standard Model and
beyond for the fundamental constituents and interactions in Nature are
demanding complex perturbative calculations involving multi-leg and
multi-loop Feynman diagrams. Currently, large effort is devoted to the
search for closed expressions of loop integrals, written whenever
possible in terms of known - often hypergeometric-type - functions. In
this work, the scalar three-point function is re-evaluated by means of
generalized hypergeometric functions of two variables. Finally, use is
made of the connection between such Appell functions and dilogarithms
coming from a previous investigation, to recover well-known results.

\end{abstract}
\vglue 2cm
\noindent{\em Keywords:} Feynman diagrams; loop integrals; hypergeometric 
series, Appell function; dilogarithm \\ \\

\vfill
 
\end{titlepage}
\subsection*{1. Introduction}
Present and foreseen unprecedented high-precision experimental results 
from $e^+e^-$ colliders
(LEP, B Factories), hadron-hadron colliders (Tevatron, LHC) and
electron-proton colliders (HERA) are demanding refined calculations
from the theoretical side as stringent tests of the Standard Model 
for the fundamental constituents and interactions in Nature.
Moreover, possible extensions beyond the Standard Model  
({\em e.g.} Supersymmetry) often require high-order
calculations where such new effects eventually would manifest.\par
In fact, much effort has been devoted so far to develop
systematic approaches to the
evaluation of complicated Feynman diagrams, looking for 
algorithms (see for example \cite{hahn,brucher} and references therein), or 
recurrence algorithms \cite{mertig,tarasov2,tarasov1}, 
to be implemented in computer program packages to cope
with the complexity of the calculation. On the other hand, the fact
that such algorithms could be based, at least in part, on already defined 
functions represents
a great advantage in many respects, as for example the knowledge
of their analytic properties ({\em e.g.} branching points and
cuts with physical significance), reduction to simpler cases with the
subsequent capability of cross-checks and so on. Commonly-used
representations for loop integrals involve special functions
like polylogarithms, generalized Clausen's functions, etc.
\par
Furthermore, over this decade the role of generalized hypergeometric functions
in several variables to express the result of multi-leg and multi-loop
integrals arising in Feynman diagrams has been widely recognized 
\cite{davy0,davy,davy2,davy3,berends,berends2}.
The special interest in using hypergeometric-type functions is twofold:
\begin{itemize}
\item Hypergeometric series are convergent within certain domains
of their arguments, 
physically related to some kinematic regions. This affords numerical 
calculations, in particular implementations as algorithms in computer
programs. Moreover, analytic continuation allows to express
hypergeometric series as functions outside those convergence domains.
\item The possibility of describing final results by means of
known functions instead of {\em ad hoc} power series is interesting by itself.
This is especially significant for hypergeometric functions because of their
deep connection with special functions whose properties
are well established in the mathematical literature. 
\end{itemize}
\par
In this paper we mainly focus on the second point, although our aim
is much more modest than searching for any master formula or method regarding 
complex Feynman diagrams. Rather we re-examine the scalar three-point 
function for massive external and internal lines, already solved in terms of 
dilogarithms \cite{olden}. The novelty of this work consists of 
solving and expressing the three-point function $C_0$, in terms of a set 
of Appell functions whose arguments are combinations
of kinematic quantities. Then, by using a theorem already proved in an 
earlier publication \cite{mas}, we end up with sixteen
dilogarithms, recovering a well-known result \cite{olden}. 
The elegant relationship shown in \cite{mas}
between Gauss and dilogarithmic functions might be a 
hint to look for more general connections
that, hopefully, could be helpful for finding compact expressions or
algorithms in Feynman calculations.

\subsection*{2. A simple relation between generalized hypergeometric series and
dilogarithms}
There are four independent kinds of Appel functions \cite{slater}, named 
$F_1,\ F_2,\ F_3,\ F_4$. In particular, the $F_3$ series is
defined as
\begin{equation}
F_3[a,a',b,b';c;x,y]\ =\ \sum_{m=0}^{\infty}\sum_{n=0}^{\infty}
\frac{(a)_m(a')_n(b)_m(b')_n}{(c)_{m+n}}\frac{x^m}{m!}\frac{y^n}{n!}
\end{equation}
which exists for all real or complex values of $a,\ a',\ b,\ b'$
and $c$ except $c$ a negative integer. With regard to its convergence,
the $F_3$ series is (absolutely) convergent when both ${\mid}x{\mid}<1$
and ${\mid}y{\mid}<1$; $(a)_m={\Gamma}(a+m)/{\Gamma}(a)$ stands for 
the Pochhammer symbol.\par
Below we re-write our main theorem published in \cite{mas}
showing a simple connection between dilogarithms \cite{lewin}
and a particular
$F_3$ Appell \vspace{0.1in} function:
\begin{equation}
\frac{1}{2}xy\ F_{3}[1,1,1,1;3;x,y]=Li_{2}(x)+Li_{2}(y)-Li_{2}(x+y-xy)
\end{equation}
\vspace{0.1in}
${\mid}\arg{(1-x)}{\mid}<\pi$, ${\mid}\arg{(1-y)}{\mid}<\pi$
and ${\mid}\arg{(1-x)(1-y)}{\mid}<\pi$. \newline
\par
Note that the validity of the theorem can be extended dropping
the restriction \linebreak[4]
${\mid}\arg{(1-x)(1-y)}{\mid}<\pi$ by slightly modifying
Eq. (2): 
\[
\frac{1}{2}xy\ F_3[1,1,1,1;3;x,y]=Li_{2}(x)+Li_{2}(y)-Li_{2}(x+y-xy)
-{\eta}(1-x,1-y)\ln{(x+y-xy)}  \nonumber
\]
where ${\mid}\arg{(1-x)}{\mid}<\pi$, ${\mid}\arg{(1-y)}{\mid}<\pi$ and 
${\eta}(a,b)=\ln{(ab)}-\ln{a}-\ln{b}$. The principal branches
of the logarithm and dilogarithm are understood.
\par
As an illustrative particular case one finds
\begin{equation}
x\ F_3[1,1,1,1;3;x,y=1]\ =\ 2x\ _3F_2(1,1,1;2,2;x)\ =\ 2\ Li_2(x)
\end{equation}
\par
Other related expressions derived from Eq. (2) may be found in \cite{mas}
\footnote{Let us note in passing a misprint in expression 3.2, Eq. (10)
of Ref. \cite{mas}: the term $\lq\lq$$-{\pi}^2/2$" has to be replaced by 
$\lq\lq$$-{\pi}^2/3$".}. Moreover, an extension of relation (3) to
polylogarithms \cite{lewin} can be easily checked
\begin{equation}
x\ F_{1:q-2;0}^{0:q;2}[\{1\}_q,\{1\}_2;\{2\}_{q-2};3;x,y=1]\ =\ 
2x\ _{q+1}F_q[\{1\}_{q+1};\{2\}_q;x]\ =\ 2\ Li_q(x)
\end{equation}
where $F_{1:q-2;0}^{0:q;2}$ stands for a kind of Kamp\'e de F\'eriet 
function \cite{kampe} (or generalized Lauricella function
of two variables \cite{srivas,exton}) defined as a series as:
\[
F_{1:t;0}^{0:r;s}[\{b\}_r,\{b'\}_s;\{d\}_t;c;x,y]\ =\ 
\sum_{m=0}^{\infty}\sum_{n=0}^{\infty}\frac{(b_1)_m...(b_r)_m
(b'_1)_n...(b'_s)_n}{(d_1)_m...(d_t)_m\ (c)_{m+n}}\ 
\frac{x^m}{m!}\frac{y^n}{n!}
\]
{\bf Proof.} It follows easily by writing the equality
\[
F_{1:q-2;0}^{0:q;2}[\{1\}_q,\{1\}_2;\{2\}_{q-2};3;x,y=1]\ =\ 
\sum_{m=o}^{\infty}\frac{\{1\}_q}{\{2\}_{q-2}(3)_m}\ \frac{x^m}{m!}\
_2F_1(1,1;3+m;1)
\]
and using the Gauss summation relation in the latter $_2F_1$ series.

\subsection*{3. Application to the evaluation of One-Loop Integrals}
In this Section, we present a simple but important application of the 
connection (2) between
hypergeometric series and dilogarithms to the evaluation
of scalar integrals appearing at one-loop level
in Feynman diagrams.
\par
In particular we shall write the scalar three-point function
corresponding to the diagram of \vspace{0.1in} Figure 1:
\par

\begin{figure}[htb]
\centerline{
\epsfig{figure=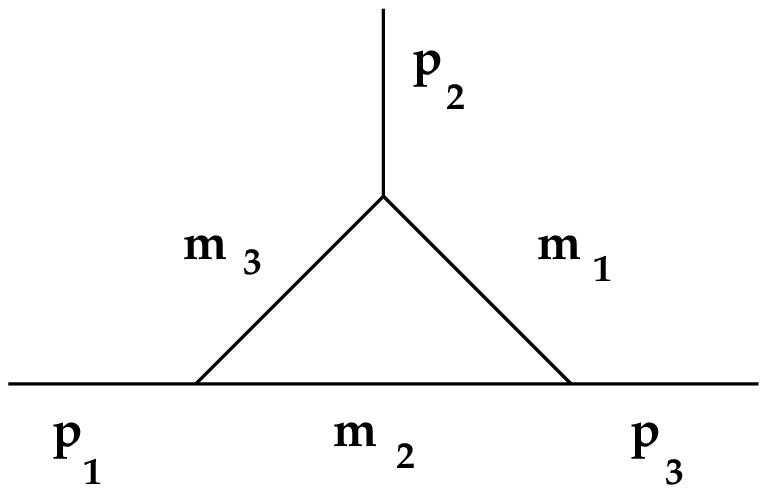,height=4.cm,width=6.0cm}}
\caption{Notation used for external and internal momenta and masses}
\end{figure}

\begin{equation}
C_0(p_1^2,p_2^2,p_3^2;m_1^2,m_2^2,m_3^2)= 
\int\ d^4q\ \frac{1}{[q^2-m_2^2][(q+p_1)^2-m_3^2]
[(q+p_1+p_2)^2-m_1^2]}
\end{equation}
\par
After Feynman parameterization \cite{thooft} one arrives at the
following double integral
\begin{equation}
C_0(p_1^2,p_2^2,p_3^2;m_1^2,m_2^2,m_3^2)/i{\pi}^2\ =\ 
\ \int_0^1u\ du\ \int_0^1 dv\ \frac{1}{Q^2}
\end{equation}
where
\begin{equation}
Q^2=
u(1-u)(1-v)p_1^2+u^2v(1-v)p_2^2+u(1-u)vp_3^2-uvm_1^2-(1-u)m_2^2-u(1-v)m_3^2
\end{equation}
\par
Let us remark that in performing the first parametric integral in Eq. (6),
either an external or an internal mass should vanish in order to get a 
hypergeometric function of the type $_2F_1(1,1;2;z)$ 
(with $z$ as a combination of masses and Feynman parameters). Specifically, 
one should drop either the $p_2^2$-term in the denominator
$Q^2$ (in this case firstly
integrating over $v$) or the $m_2^2$-term (in this case
firstly integrating over $u$), 
of course the particular choice of index 2 by no means representing 
any loss of generality.\par 
Hence, the basic idea is to follow a procedure to get rid of any of those 
masses, so that one can express the final result in terms of generalized
Gauss functions in two variables, after performing  the 
parametric integrations over $v$ and $u$ consecutively \cite{mas2}. 
In fact, this goal can be achieved in a standard
way by means of a $\lq\lq$trick" introduced 
by Passarino and Veltman \cite{passa} in this context, using a propagator 
identity and defining a set of unphysical vectors 
\cite{thooft,mas2}. Then, the loop integral (5) can be
splitted into two pieces corresponding to the sum
of two three-point functions with either an internal or an
external mass equal to zero. The latter possibility may be
exploited for one (at least) timelike external momentum - in fact this is 
generally the case at next-to-leading order calculations
since for any real scattering or decay process there is 
always an external timelike or lightlike momentum. \par
Therefore, under the assumption of any timelike external momentum
we write below the final
expression of the three-point function for massive internal and
external lines in a very closed form:

\begin{eqnarray}
C_0(p_1^2,p_2^2,p_3^2;m_1^2,m_2^2,m_3^2)/i{\pi}^2 & = & 
\frac{{\sigma}_{\alpha}}{2\lambda^{1/2}(p_1^2,p_2^2,p_3^2)}\ {\times}  \\
& & \sum_{i=1}^{2}\ \biggl(\ [\ R(x_i,y)-R(x_i',y)\ ]\ -\ 
[\ \alpha{\leftrightarrow}(1-\alpha), 1{\leftrightarrow}3\ ]\ \biggr)
\nonumber
\end{eqnarray}
where the interchange of indices $1{\leftrightarrow}3$ does not
apply inside $\alpha$. The notation is the \vspace{0.1in} following:
\par
\begin{itemize}
\item Each term of the summation (we use the same symbol $R$ as in \cite{olden})
can be written in terms of an Appell's function according to
\[ R(x_i,y)\ =\ \frac{1}{2}\ x_iy\ F_3[1,1,1,1;3;x_i,y] \]
\item ${\alpha}$ is given by any of the two solutions:
\[ {\alpha}_{\pm}\ =\ 
\frac{p_1^2+p_2^2-p_3^2\ {\pm}\ {\lambda}^{1/2}(p_1^2,p_2^2,p_3^2)}
{2p_2^2} \]
with ${\lambda}$ standing for the K\"{a}llen function, 
$\lambda(x,y,z)= (x-y-z)^2-4yz$; ${\sigma}_{\alpha}={\mp}1$ according
to the choice for the square root of ${\alpha}_{\pm}$. 
\par
It is not difficult to verify the independence of the result as
expressed in Eq. (8) on the choice of the sign for $\alpha$
using the initial symmetry of the scalar integral under the permutation of
indices, in particular $1{\leftrightarrow}3$ (now including $\alpha$ 
under such index permutation). Hence the first four $R$ terms in (8)
would depend on $\alpha(3,1)$, instead of $\alpha(1,3)$. Since
$\alpha_+(3,1)=1-\alpha_-(1,3)$) one arrives quickly to the above-mentioned
conclusion taking into account $\sigma_{\alpha}$.
\item Variables $x_i$ and $x_i'$ is defined as: $x_i=1/r_i$
and $x_i'={\alpha}/t_i$ where
\[ r_{1,2}\ =\ \frac{p_1^2+m_3^2-m_2^2\ {\pm}\ 
{\lambda}^{1/2}(p_1^2,m_3^2,m_2^2)}{2p_1^2} \]
and
\[ t_{1,2}\ =\ \frac{p_2^2+m_3^2-m_1^2\ {\pm}\ 
{\lambda}^{1/2}(p_2^2,m_3^2,m_1^2)}{2p_2^2} \]
\item The $y$ variable is given by
\[ y\ =\ \frac{\alpha\ {\lambda}^{1/2}(p_1^2,p_2^2,p_3^2)}
{(1-\alpha)[-p_1^2+m_3^2-m_2^2]+\alpha[-p_3^2+m_1^2-m_2^2]} \]
\end{itemize}
\vskip 0.5 cm
\par

Now, using Eq. (2) the scalar one-loop integral of Figure 1 can be 
directly expressed in
terms of sixteen dilogarithms since there is a pairwise cancellation
between eight $Li_2(y)$ functions, leading to a similar
result 
\footnote{Let us note, however, some discrepancies between both expressions
once notations are made equivalent.
According to us there is a misprint in the
definition of their $w_3^{\pm}$ variables in Eq. (50) of Ref. \cite{olden}. 
This question was already discussed by Oldenborgh and M.A.S.L. time ago.}
as obtained in Ref. \cite{olden}. Therefore, it becomes
apparent the usefulness of the connection between dilogarithms and
generalized hypergeometric functions shown in this theorem \cite{mas}.
Expectedly, further relations involving multiple hypergeometric
functions and  polylogarithms or related functions \cite{kolbig} 
could be discovered \cite{mas3}, with the possibility of application to
more complicated loop integrals.\par
Finally, for the sake of completeness we show in Figure 2
some (finite) results considering several special values for the 
internal and external masses, 
as limits deriving from the main expression \vspace{0.1in} (8).
\par

\begin{figure}[htb]
\centerline{
\epsfig{figure=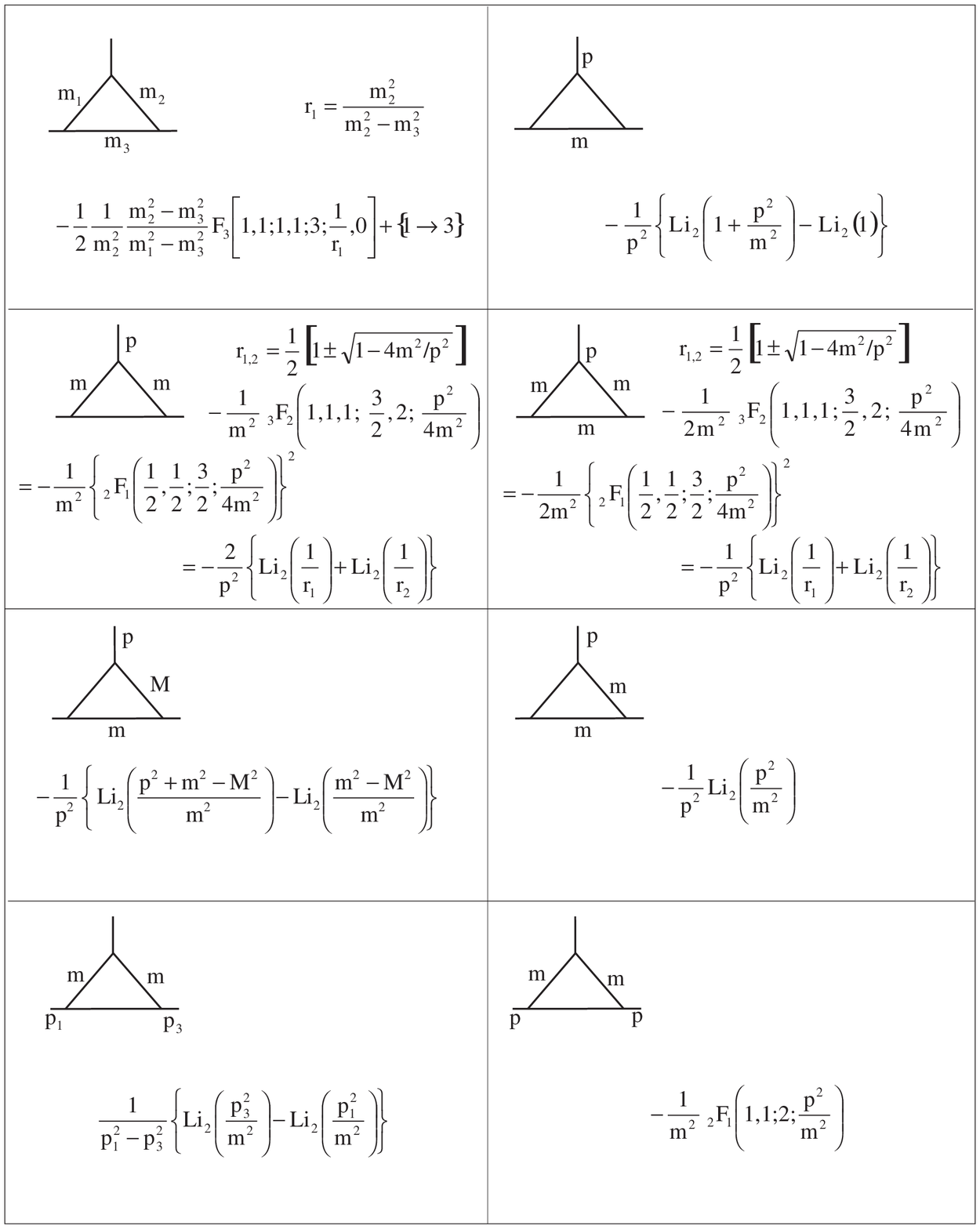,height=18.cm,width=14.5cm}}
\caption[Table 1]{Special finite cases for the $C_0/i{\pi}^2$ three-point 
function.
Missing symbols on external and internal lines mean zero mass.}
\end{figure}

\subsection*{4. Summary and last remarks}
In this paper, we have shown that the familiar scalar three-point function 
$C_0$ arising in Feynman diagram calculations, can be expressed in terms
of eight Appell functions
whose arguments are simple combinations of internal and
external masses. The extension to complex values
of the masses (of concern  in several interesting physical cases)
can be performed via 
analytic continuation of the hypergeometric functions. Moreover, by
invoking a theorem proved in \cite{mas} connecting
dilogarithms and generalized hypergeometric functions, we have 
recovered a well-known formula for the evaluation of $C_0$
\cite{olden}. Finally, let us stress that, although its simplicity, 
this work is on the line of searching for closed expressions 
of Feynman loop integrals. On the other hand but in
a complementary way, such physical requirements should motivate, on the 
mathematical side, further developments of the still largely unknown
field on multiple hypergeometric functions (as compared to
one-variable hypergeometric functions) and their connections to
generalized polylogarithms \cite{kolbig}.

\subsubsection*{Acknowledgments}
This work has been partially supported by CICYT under grant
AEN-96/1718. M.A.S.L. acknowledges an Acci\'on Especial (code 6223080)
by the regional government Generalitat Valenciana. L.G.C.R. has been 
supported by a fellowship from the D.G.A.P.A.-U.N.A.M. (M\'exico).


\end{document}